\documentclass[11pt]{article}

\usepackage{lipsum} 

\usepackage{bm}
\pdfoutput=1
\usepackage{amsmath,amsfonts,amsthm}
\usepackage{relsize}
\usepackage{esdiff}  
\usepackage{booktabs}  
\usepackage{url}  
\usepackage{cleveref}  
	\crefname{equation}{equation}{equations}
	\crefname{figure}{figure}{figures}	
	\crefname{table}{table}{tables}
\usepackage[skip=1.5pt,font=small]{caption}

\usepackage{caption}
\usepackage{bbm}

\usepackage[usenames,dvipsnames,svgnames,table]{xcolor}

\usepackage{graphicx}
\usepackage{hyperref}  

\usepackage[sc]{mathpazo} 
\usepackage[T1]{fontenc} 
\usepackage{microtype} 

\usepackage{multicol} 
\usepackage[margin={1cm,1.5cm}]{geometry}
\usepackage{booktabs} 
\usepackage{float} 
\usepackage{hyperref} 

\usepackage{lettrine} 
\usepackage{paralist} 

\usepackage{abstract} 
\usepackage{abstract} 

\usepackage{titlesec} 
\renewcommand\thesection{\Roman{section}} 
\renewcommand\thesubsection{\Alph{subsection}} 
\titleformat{\section}[block]{\large\scshape\centering\bfseries}{\thesection.}{1em}{} 

\titleformat{\subsection}[block]{\scshape\centering}{\thesubsection.}{1em}{} 

\usepackage{fancyhdr} 
\DeclareCaptionFormat{myformat}{#1#2#3\hrulefill}
\usepackage{authblk}
\makeatletter
\renewcommand\AB@affilsepx{, \protect\Affilfont}
\makeatother
\title{\vspace{-15mm}\fontsize{16pt}{16pt}\selectfont\textbf{A Neural Networks Approach to Predicting How Things Might Have Turned Out Had I Mustered the Nerve to Ask Barry Cottonfield to the Junior Prom Back in 1997}} %
\author[1]{Eve Armstrong\thanks{earmstrong@ucsd.edu}}
\affil[1]{BioCircuits Institute, University of California, San Diego, La Jolla, CA 92093-0374}
\date{(Dated: April 1, 2017)}
\renewcommand\Affilfont{\itshape\small}
\begin{document}
\maketitle 
\begin{abstract}
\vspace{6mm}
We use a feed-forward artificial neural network with back-propagation through a single hidden layer to predict Barry Cottonfield's likely reply to this author's invitation to the \lq\lq \textit{Once Upon a Daydream}\lq\lq\ junior prom at the Conard High School gymnasium back in 1997.  To examine the network's ability to generalize to such a situation beyond specific training scenarios, we use a L2 regularization term in the cost function and examine performance over a range of regularization strengths.  In addition, we examine the nonsensical decision-making strategies that emerge in Barry at times when he has recently engaged in a fight with his annoying kid sister Janice.  To simulate Barry's inability to learn efficiently from large mistakes (an observation well documented by his algebra teacher during sophomore year), we choose a simple quadratic form for the cost function, so that the weight update magnitude is not necessary correlated with the magnitude of output error. 

Network performance on test data indicates that this author would have received an 87.2 (1)\% chance of \lq\lq Yes\rq\rq given a particular set of environmental input parameters.  Most critically, the optimal method of question delivery is found to be Secret Note rather than Verbal Speech.  There also exists mild evidence that wearing a burgundy mini-dress might have helped.  The network performs comparably for all values of regularization strength, which suggests that the nature of noise in a high school hallway during passing time does not affect much of anything.  We comment on possible biases inherent in the output, implications regarding the functionality of a real biological network, and future directions.  Over-training is also discussed, although the linear algebra teacher assures us that in Barry's case this is not possible.  
\end{abstract}
\section{INTRODUCTION}
\begin{multicols}{2}
By most historical accounts, Barry Cottonfield was a brash, unprincipled individual (private communications 1983-1998; Lunch Ladies' Assoc. v. Cottonfield 1995) with slightly below-average learning capabilities (Consolidated Report Cards: 1986-1998), and with a reputation among the local police force for an alleged - yet unprovable - association with the serial disappearance of various neighbors' cats (WHPD Records Division 1988, 1989a, 1989b, 1996).  

But they didn't know him like I did.  The moment I first laid eyes on Barry - across the see-saw in second grade - I imagined that he was deeply multi-layered: a mysterious soul with a hidden strength driven by latent, and perhaps unstable, inclinations.  I could tell.  It was sure to be reflected somewhere in those expressive eyes of his, if one were to peer hard enough (see Fig. 1).  Further, this sentiment was in-part justified when he left me dangling up on the see-saw for several minutes until the playground monitor intervened.  As noted, not many others saw any nuanced side of Barry.  This was due to the fact that one had to exert enormous effort in order to see it.  Obsessive, in fact. 

Throughout our four years at Conard High School in West Hartford, Connecticut, Barry and I engaged in numerous in-depth conversations - usually in the hallway during the four-minute permitted passing time.  During these opportune intervals I peppered him with questions regarding himself: habits, quirks, predilections, revul- 
\begin{figure}[H]
\centering
  \includegraphics[width=80mm]{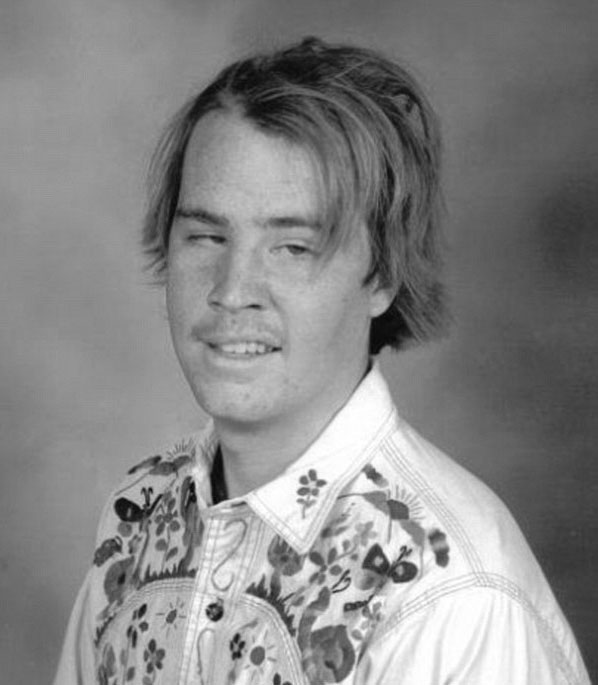}    
  \caption{\textbf{Barry Cottonfield in 1997} (\textit{Yearbook Pictures}).}
\end{figure}
\noindent
sions, sunniest plans, bloodiest secrets, and his essential identity.  I took copious notes on his responses, whenever he was willing to offer me any.  I recorded them instantly and later transferred them to a diary (well, diarie\textit{s} - ultimately five volumes).  The frequency of my entries tapered somewhat toward the end of senior year, by which time Barry had learned that he could maximize the chances of evading me by ducking out of class two minutes early (note that he required 3.5 years to learn this; see \textit{Results: Over-training}).  Catching up to him often required an all-out sprint on my part.  But it was worth it.  He was fascinating.  

I never, however, mustered the courage to ask Barry to the prom junior year.  This decision has haunted me; I wonder what his answer would have been.  While an unparalleled number of poems has been written on this topic (Armstrong 1998-2017) and hundreds of sleep hours wasted on \textit{what-if?} dreams, no quantitative study has ever been undertaken to investigate the question.  Recently I decided to address this situation.  I wanted an answer - or at least some maximum-likelihood answer.

But how to obtain an answer?  Contacting Barry to simply pose the question was no longer a possibility; it would have violated my restraining order (WHPD Records Division 2002-2017, renewable).  I had, however, saved my copious notes on our question-and-answer sessions - notes that might serve as a proxy of Barry's personality.  Might I harness this information to somehow reconstruct Barry - and use that representation to \textit{predict backwards}?  To investigate this possibility, let us now turn to the black art of artificial neural network construction.

Artificial neural networks (ANNs) are a type of machine learning algorithm in which a neurobiology-inspired architecture is created, via exposure to training data, to mimic the capability of a real brain to categorize information (e.g. Hopfield 1988, Jain \& Mao 1996).  This artificial brain, while astoundingly stupid by human standards (and by fruit-fly and sea-slug standards, incidentally), can nevertheless learn to be stupid rather quickly and thereafter serve as a powerful tool for particular predictive purposes.  

The combined features of relatively low intelligence and limited utility parallel Barry Cottonfield quite readily.  This framework, then, appears well-suited for creating an artificial Barry and examining its decision-making strategies.  Decision trees and decision-tree/ANN hybrids (e.g. Lee \& Yen 2002, Biau et al. 2016, Rota Bulo \& Kondschieder 2014, Kontschieder et al. 2015) may also mimic such calculations, but for our purposes it seems important to constrain the algorithm to a brain-like framework.  Specifically, we aim to recreate the brain that was Barry's at the time immediately preceding the junior prom in May 1997.  The five-volume high school diary can be employed as a training set.

In this paper we examine a two-tiered learning approach to infer the likely outcome of inviting Barry Cottonfield to the junior prom twenty years ago.  In the first tier, a zeroth-order Barry is set based on known answers to simple questions of identity and person tastes, which are taken to be independent of immediate environment.  In the second set, zeroth-order Barry receives questions that require a decision to be made; these decisions might be environment dependent.  In this phase, zeroth-order Barry is tweaked to account for environmental preferences, and a corresponding optimal set of environmental parameter values is identified.  Finally, the prediction phase employs this optimal set to yield an 87.2 (1)\% likelihood of \lq\lq Yes\rq\rq\ to the prom invitation.  

While the output of ANNs does not necessarily provide the right answer, we find ourselves quite enamored of this one.  Consequently we take it to be right.
\end{multicols}
\newpage
\section{LEARNING ALGORITHM}
\begin{multicols}{2}
\subsection{\textbf{Framework}}

The architecture consists of one input, one hidden, and one output layer (Fig. 2).  The number of input nodes is 137.  The inputs are: 1) a question, and 2) a set of 136 environmental parameters.  The output is binary (Yes or No).  To maximize the network's ability to generalize beyond the training set, we choose the number of hidden nodes to be 1,000.  Now, a larger-than-necessary hidden layer may increase computational demands and the likelihood of overfitting.  These are not concerns for our purposes, however, as we have no time constraints and it is highly improbable that a realistic model of Barry can be over-trained (Consolidated Report Cards: 1986-1998).

The input signal propagates forward only.  Synaptic weights and node biases are updated via the gradient-descent minimization of a cost function that represents output error.  Updates are effected via a constant back-propagation rule that relates output error to the cost function derivative with respect to the weights and biases (see \textit{Materials and Methods}).  
\begin{figure}[H]
\centering
  \includegraphics[width=55mm]{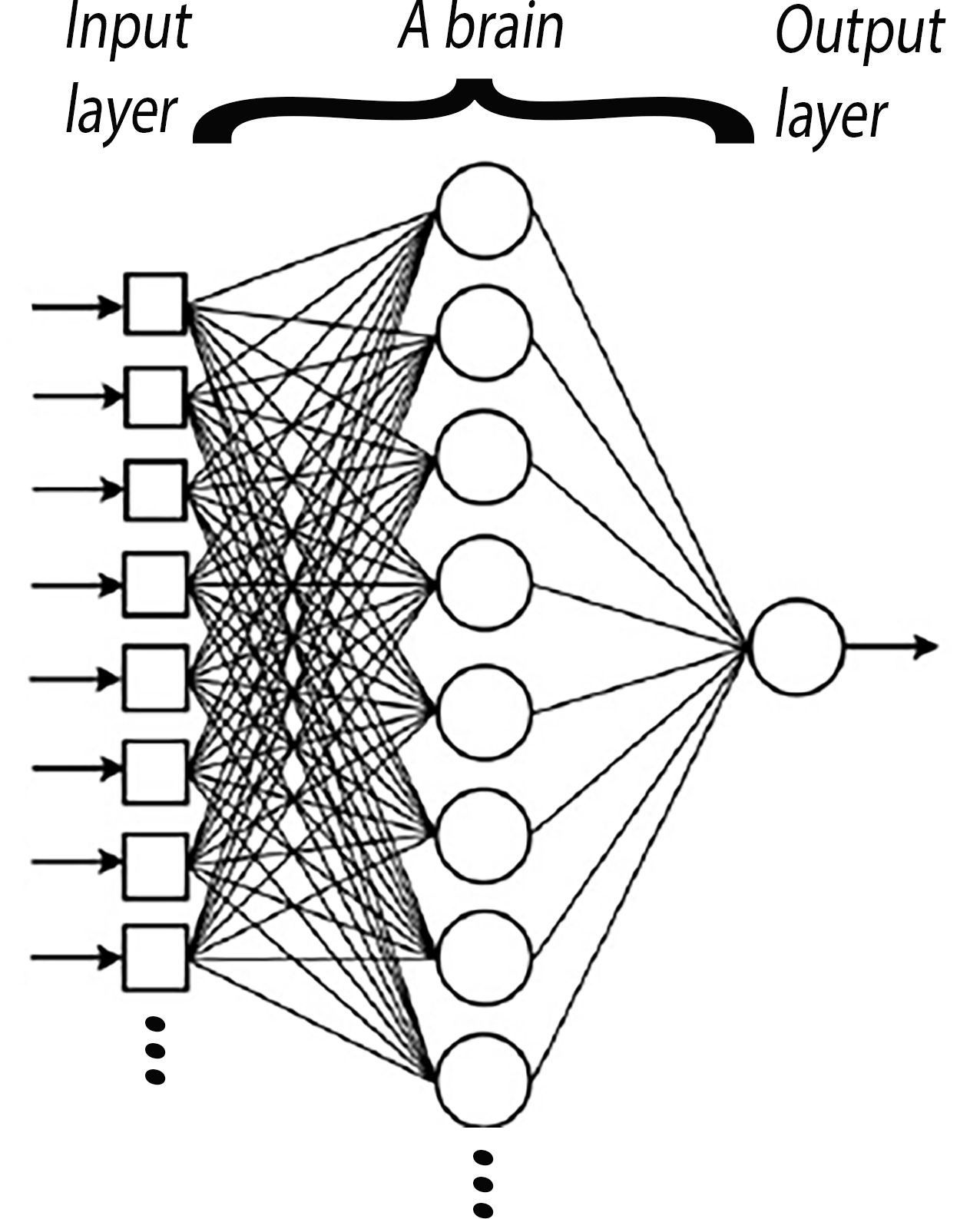}
\vspace{3mm} 
  \caption{\textbf{Basic brain architecture to be trained.}  There is one input, one hidden, and one output layer.  The number of input nodes is 137.  The inputs are: 1) a question, and 2) a set of 136 environmental parameters.  The output is binary (Yes or No).  To maximize the network's ability to generalize beyond the training set, the number of hidden nodes is chosen to be 1,000.  While a larger-than-necessary hidden layer may increase computational demands and the likelihood of overfitting, these are not concerns for our purposes.  First: we have plenty of time, and second: it is highly improbable that a realistic model of Barry can be over-trained (Consolidated Report Cards: 1986-1998).}
\end{figure}

Now, this choice of cost function implicitly takes Barry's aim in life to be the minimization of error.  This assumption is not squarely consistent with his behavior in a classroom setting (see \textit{Discussion}).  To mitigate the possible impact of this bias, we choose a simple quadratic term for this cost function, so that the weight update magnitude does not necessary correlate with the magnitude of output error.  This choice aims to simulate Barry's inability to learn efficiently from large mistakes (Report Card: Algebra 1996).  Finally, we include a regularization term to examine the network's sensitivity to noise (see \textit{Materials and Methods}).  When regularization strength is chosen aptly, this term penalizes high synaptic weights - and low weights may enable a model to more readily recognize repeating patterns, rather than noise, in a data set (Nielsen 2015).  

The algorithm converges upon a cost value that is not guaranteed - nay, is highly unlikely - to be the global minimum.  It will be sufficient for our standards, however, provided that our standards are sufficiently low.

\subsection{\textbf{Two-stage learning}}

Learning occurs in two stages, denoted: \textit{Becoming Barry} and \textit{Tweaking Barry}, to be described below.  Following each stage is a validation and a test step.  In each stage, a single data set is used, from which we select random batches for training, validation, and testing.  The percent accuracy is measured continually, and learning is stopped when that percentage begins to asymptote to a stable value.  Note that as the questions were originally posed in a high school hallway during passing time, these training sets provide an opportunity to probe the network's performance in a high-noise environment.  

The use of a single data set from which to draw training, validation, and test questions might reduce Barry's ability to generalize to broad situations.  We do not consider this a problem, however.  The inclusion of a regularization term and an absurdly large number of hidden nodes is intended to foster generalizability.  Moreover, the prediction phase involves a single question that is typical of the scenario at hand.  That is: for the purposes of this study, all Barry needs to do is learn how to be a high school junior.\\
\newpage
\noindent
\textit{Stage 1: Becoming Barry}

In this first stage of learning, the architecture adjusts to a zeroth-order representation of Barry's brain as it existed in 1997, given a set of questions that are taken to be independent of immediate environment.  These questions probe knowledge of Barry's basic identity, personal tastes, interests, and emotional tendencies.  There are 10,002 entries in the question set.  Samples, including: \textit{Is your name Barry?}, are listed in Table 1.  It is assumed (well, hoped) that these questions require no decision-making by the time Barry has reached age 17.

Note that while the network contains 137 input nodes, only Input Node 1 receives a signal during this stage - the signal being a question.  The other 136 nodes exist to receive environmental input parameters, and we set these aside for the moment.  In this first stage, then, the only weights that are updated are those associated with Input Node 1.  These consist of the input-to-hidden-layer weights that emanate from Input Node 1 only, and all of the hidden-to-output weights.  Fig. 3 (\textit{left}) denotes these weights in purple.  Note that while node bias is not depicted in Fig. 3, all nodes receiving input will receive bias updates.    
\end{multicols}
\begin{figure}[H]
\centering
  \includegraphics[width=82mm]{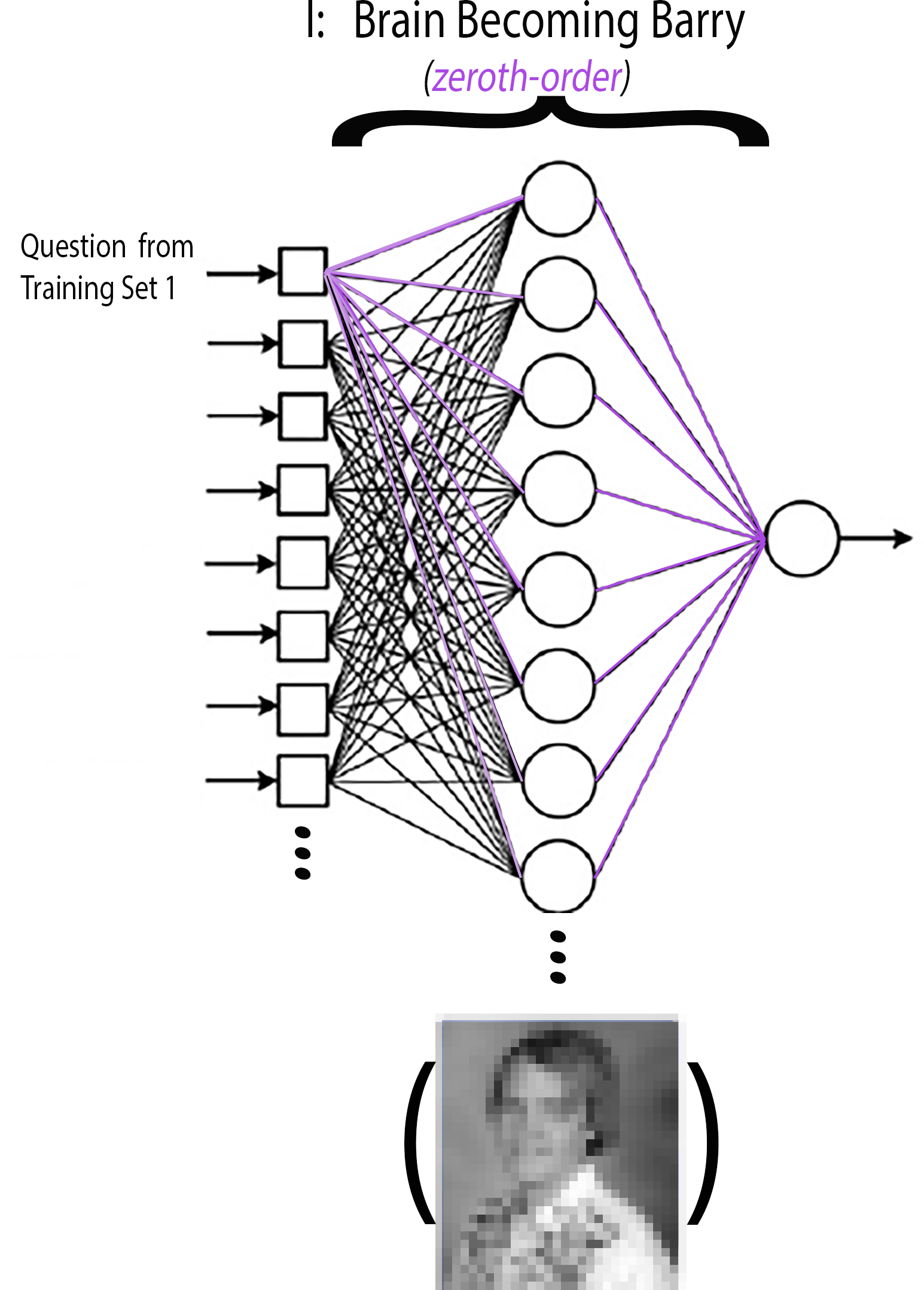}
  \hspace{2cm}
  \includegraphics[width=82mm]{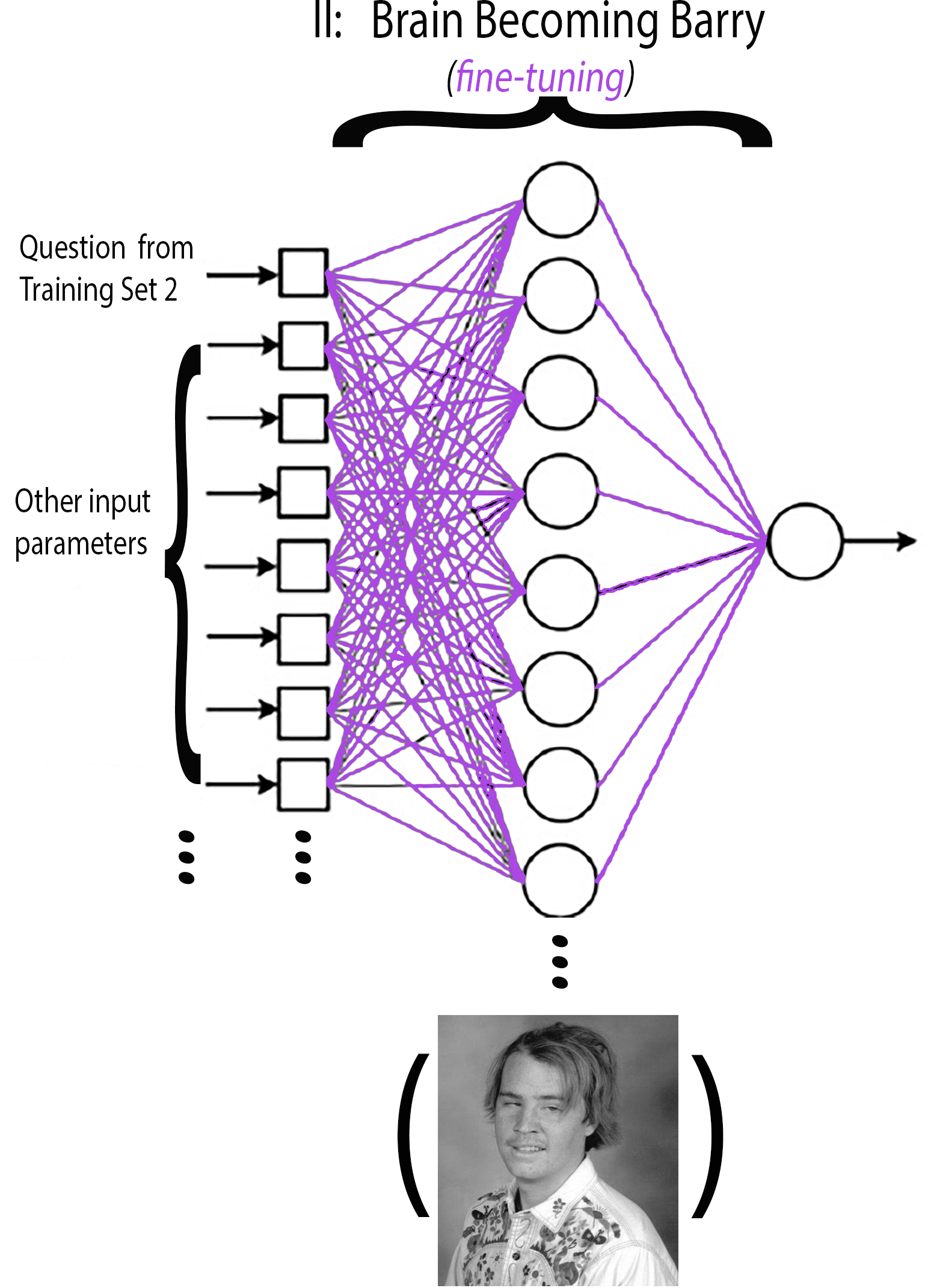}\\  
  \caption{\textbf{Two-stage learning procedure.}  \textit{Left}: Setting zeroth-order Barry.  Here we update weights and biases (biases are not pictured) based on simple questions regarding basic identity, which we take to be environment-independent.  \textit{Right}: Here we tweak zeroth-order Barry using questions that require contemplation or decision-making.  These questions may depend in part on environmental parameters.  We perform this step multiple times on all possible sets of environmental parameter values.  To identify Barry's preferred set, we assume that some information regarding Barry's environmental preferences have been implicitly wired into his zeroth-order personality.  Thus, we take as optimal the set requiring the \textit{least} tweaking of the zeroth-order weights that were set during Stage 1.}
\end{figure}
\begin{multicols}{2}
\noindent
\textit{Stage 2: Tweaking Barry}\\
\indent Having constructed a zeroth-order Barry, we now expose him to questions that require a decision to be made, where now the answer may depend on particular environmental parameters.   

We consider 136 parameters that, given this author's conversational history with Barry, are suspected to affect Barry's response.  These include, for example: \textit{Time of Day}, \textit{Question Delivery Method} (e.g. Secret Note versus Verbal Speech), and \textit{Type of Outfit the Question-deliverer is Wearing}.  We would like to ascertain Barry's preferences regarding the parameter values.  In this arrangement, Input Node 1 still receives each question, and Input Nodes 2-137 now receive the environmental input signals, which are elements of a vector $\bm{p}$.

As in Stage 1, the questions and answers for Stage 2 are taken from this author's high school diary.  Further, a significant fraction of the questions were asked multiple times over various environmental contexts; this point shall prove pertinent momentarily.  Sample questions are listed in Table 2. 

Now, as Barry's basic identity has been programmed in Stage 1, we imagine that some sense of environmental preferences are now implicit in his personality.  For example, consider this question from Set 2: \textit{Can I borrow five dollars?.}   During Stage 1 we had wired in the answer to \textit{Is your favorite color beige?} as Yes.  It is possible that that answer contains information regarding the likelihood that Barry will lend us five dollars if we are wearing a beige sweater at the time.  One environmental parameter to be tuned, then, is Outfit Color.  

What is the mapping between Barry's intrinsic wiring from Stage 1 and parameter preferences $\bm{p}$ in Stage 2?  That is, what is the set of functions $\bm{f}$ that takes the matrix $\bm{Barry_{zero}}$ such that: $\bm{p} = \bm{f}(\bm{Barry_{zero}})$?  If we knew $\bm{f}$, we would be able to infer the optimal set $\bm{p}$.  We do not know $\bm{f}$, however, and we will not even attempt to construct it.  Instead, we will assume that within zeroth-order Barry there exists already an implicit approximate preference for set $\bm{p}$, and thus that the optimal choice of $\bm{p}$ should affect zeroth-order Barry minimally. 

Our aim in Learning Stage 2, then, is to seek the set of environmental parameter values $\bm{p}$ that requires the \textit{least tweaking} of the synapse weights that were estimated in Stage 1, assuming that this set will be the most faithful representation of Barry's true preferences.  Fig. 3 (\textit{right}) shows in purple the weights that are now to be estimated: the weights estimated in Stage 1 in addition to the weights associated with the input nodes that receive the 136 environmental signals.

For each question in Set 2, we train Barry independently on all possible combinations of environmental parameter values $\bm{p}$.  We choose the optimal set regardless of the required training duration or the percent accuracy of the output.  

Now, the optimal set $\bm{p}$ should be independent of the question posed, with one salient exception.  As noted earlier, this author asked many of the questions in Set 2 on multiple occasions, in various environmental contexts.  Answers were found to be stable across all parameter values, except one: instances in which Barry had recently engaged in a fight with his annoying kid sister Janice (Fig. 4).  Given this observation, we perform Learning Stage 2 \textit{twice}: once for the scenario in which a fight with Janice had recently occurred, and one for the scenario in which it had not.  Each version corresponds to a distinct training set.

The Stage 2 training begins with the weights and biases estimated in Stage 1 at those starting values, and with all as-yet unestimated weights and biases initialized randomly.  Finally, it is assumed that since this author is the individual who posed all 24,203 questions present in the training data (see \textit{Materials and Methods}), Barry has been conditioned to assume that any question posed is coming from this author.  That is: during the prediction phase, he will know who is asking him to the prom. 
\begin{figure}[H]
\centering
  \includegraphics[width=70mm]{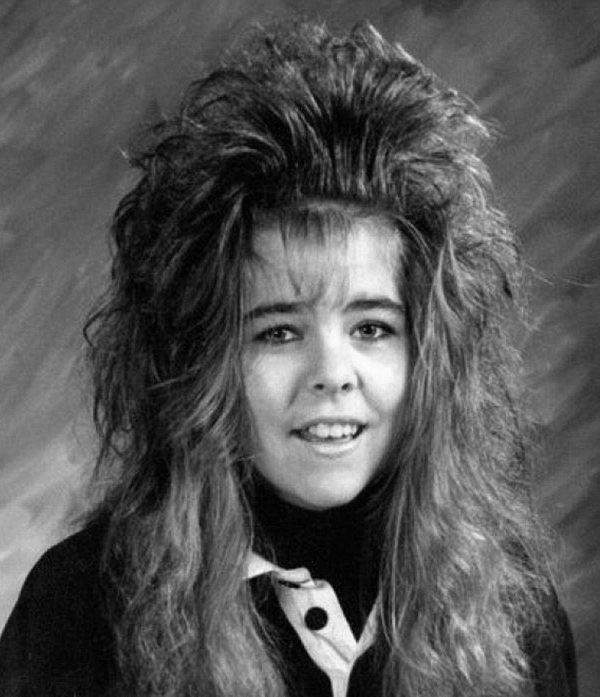}
  \caption{\textbf{Janice Cottonfield, Barry's kid sister, in 1997} (\textit{Yearbook Pictures}).}
\end{figure}

\end{multicols}
\setlength{\tabcolsep}{1pt}
\begin{table}[H]
\small
\centering
\caption{\textbf{Samples from the 10,002-item Training Set 1: \textit{Becoming Barry}}} 
\begin{tabular}{ l c|c c} \toprule
 \textit{Question re: Identity} & Answer & \textit{Question re: Personal inclinations} & Answer \\\midrule  
 \textit{Is your name Barry?} & Yes & \textit{Do you wish your name were Jackson?} & No \\
 \textit{Do you have a dachshund named Raskolnikov?} & Yes & \textit{Was \lq\lq Raskolnikov\lq\lq\ your idea?} & Yes \\
 \textit{Do you own a peanut plant?} & Yes & \textit{Do you like peanuts?} & No \\  
 \textit{Do you play lacrosse?} & Yes & \textit{Are you the best lacrosse player ever?} & Yes \\ 
 \textit{Is your house beige?} & No & \textit{Is your favorite color beige?} & Yes \\
 \textit{Do you have ten toes?} & Yes & \textit{Would you like to have ten toes tomorrow?} & Yes \\ \\\bottomrule
\end{tabular}
\newline
\small{These questions concern information about basic identity and personal tastes; they are designed to set zeroth-order Barry.  Answers are taken to be independent of environment.} 
\end{table}
\begin{multicols}{2}

\end{multicols}
\setlength{\tabcolsep}{1pt}
\begin{table}[H]
\small
\centering
\caption{\textbf{Samples from Training Set 2: \textit{Tweaking Barry}; answer depending on occurrence of a recent fight with Janice}} 
\begin{tabular}{ l|c c} \toprule
 \textit{Question} & Answer &  \\
 \textit  &  (Janice-fight: Yes) & (Janice-fight: No) \\\midrule  
 \textit{Can I borrow a dollar?} & No & No \\
 \textit{Can I borrow five dollars?} & No & No \\
 \textit{Is this a good time?} & No & No \\
 \textit{Are you going to hand in Mrs. Lindsey's extra credit assignment?} & Yes & No \\
 \textit{Will you look after the school mascot ferret over this weekend?} & Yes & No \\   
 \textit{Do you want to buy a kitten?} & Yes & No\\
 \textit{Can you keep a secret?} & No & No \\
 \textit{Please stop clicking your tongue behind me all through chem class?} & Yes & No\\    
 \textit{Would you consider going to the prom with Mindy?} & Yes & Yes\\
 \textit{What about Lydia?} & Yes & No \\ \\\bottomrule
\end{tabular}
\newline
\small{These questions are designed to probe Barry's decision-making strategies given various environmental parameters.  They were asked under two conditions: once at times when Barry had recently fought with his kid sister Janice, and once at times when he had not.  The training sets for \lq\lq Janice-Fight: Yes\rq\rq\ and \lq\lq Janice-Fight: No\rq\rq\ contain 7,345 and 9,557 questions, respectively.  2,701 questions overlap between scenarios.  The samples above are drawn from the \textasciitilde\ 50\% of these 2,701 overlapping questions for which Barry gave a reliable answer in the \lq\lq Janice-fight: Yes\rq\rq\ case.  For the other 50\%, Barry's answer failed to stabililize, suggesting that fighting with his sister rendered him quite unlike himself (see \textit{Results}).}
\end{table}
\section{RESULTS}
\begin{multicols}{2}
\subsection{\textbf{Using the \textit{No-fight-with-Janice} results to obtain predictions}}

We chose to make predictions using results of the No-Fight-With-Janice training set.  This decision was made for two related reasons: with this set, we obtained 1) less updating of Barry's zeroth-order synaptic weights and 2) greater stability of the output answer.

Regarding the first point: the set for which the answer to \textit{Have you recently fought with Janice?} is Yes modified the weights of Stage 1 by roughly eight times as much as did the set for which the answer is No.  This suggests that fighting with Janice made Barry sufficiently vexed that he was significantly not himself anymore.

Secondly: on many trials of \lq\lq Janice-fight: Yes\rq\rq, learning of a significant fraction of the questions did not asymptote to a stable answer.  Specifically: on 76\% of trials, at least 50\% of the questions failed to converge during the time required for the other fraction of questions to attain stable values and for the percent accuracy of those values to asymptote.  For questions that did asymptote to a stable value, the typical required duration for learning, over all trials, was 31 (4) minutes.  This lower limit is consistent with the observation that the effects of fights with Barry's sister tended to last at least 45 minutes (Consolidated Guidance Counselors' Records 1995-1997).  Furthermore, for the case in which a recent fight with Janice had occurred, the back propagation resulted in seemingly random redistributions of weights.  This phenomenon appears to be consistent with Barry's deviation from zeroth order, as noted above; namely: the random weight changes suggest that Barry was forgetting himself entirely.  We do note, however, that for some brains, random changing of synaptic weights can be just as fine a paradigm for learning as is a rigid back-propagation rule (Lillicrap et al. 2016).

One final reason we took to dismiss the results of the \lq\lq Janice-fight: Yes\rq\rq\ scenario was a simple examination of the responses that \textit{did} asymptote to stable values.  Within this category, for any questions that represented \textit{requests for favors}, Barry was 27 (8)\% less likely to acquiesce following a fight.  To make predictions, then, we adopted the optimal parameter set that emerged from the training scenario in which Barry had \textit{not} recently fought with Janice.  

For this optimal set, there emerged one parameter that was consistently (34 (4)\%) more likely to yield a favorable reply to a request.  This parameter was \textit{Question Delivery Method}: Secret Note is favored over Verbal Speech.  Burgundy Mini-dress won out slightly (at the 2 (1)\% level) to other shades of mini-dress.  The Mini-dress category overall, meanwhile, beat 726 other wardrobe choices at the 6 (2)\% level.  The remaining 134 environmental parameter choices had negligible effect on outcome.

\subsection{\textbf{Prediction}}

Given the optimal set of parameter values in a scenario in which Barry had not recently engaged in a fight with Janice, and over 1,000 trial predictions, the prom invitation yielded an 87.2 (1)\% success rate.

\subsection{\textbf{Over-training}}

As noted, most questions were originally delivered in a high school hallway, which is typically an extremely noisy environment.  It is possible, then, that answers Barry gave during Stage 2 were influenced by factors that were not accounted for in the environmental parameter set.  To examine the model's sensitivity to noise, we examined results for all choices of regularization strength.  Regularized networks are constrained to build simple models based on patterns seen often in training, whereas unregularized networks may be more likely to learn the noise.

For all choices of regularization strength, the network performed comparably.  This suggests that the nature of noise in a high school hallway during passing time does not affect much of anything.  Alternatively, Barry simply did not hear it, which is the same conclusion stated differently.  Moreover, either way, it seems to confirm the algebra teacher's report that to over-train Barry would require a monumental undertaking.

\subsection{\textbf{Inability to learn particular questions}}

We make a final observation that pertains again to the algebra teacher's comment on over-training.  Even for the training set corresponding to the \lq\lq Janice-fight: No\rq\rq\ scenario, we identified a small subset of questions (2\% of the sample) for which Barry's response never attained a stable value.  In particular, Barry learned quickly that he had ten toes but never a reliable response to \textit{Will you have ten toes tomorrow?}, despite trials over consecutive days.  This finding might be indicative of one problem inherent in attempting to teach a memory-less system to learn something (see \textit{Discussion}.)
\end{multicols}
\newpage
\section{DISCUSSION}
\begin{multicols}{2}
\subsection{\textbf{A memory-less system that learns?}}

We comment on Barry's inability to yield a stable output to a small subset of questions from Stage 2: \textit{Tweaking Barry}, even in the regime in which he had not recently fought with Janice.  These questions shared a commonality: they required imagining the future.  One question, for example, was: \textit{Will you have ten toes tomorrow?}.  The answer failed to stabilize despite training sessions over consecutive days - perhaps reflecting one of the inherent difficulties of teaching memory-less systems to learn.  

It is possible that this finding provides further evidence in favor of his teacher's theory that Barry cannot be over-trained.  Alternatively, however, perhaps our experimental design \textit{underestimated} Barry's learning capacity.  Indeed, while it is highly likely that Barry will have ten toes tomorrow (given ten toes today), it is not guaranteed.  Barry may have apprehended this uncertainty.  The capacity for such aptitude was not captured in the experimental design, an omission that may underlie the instability of the output.
\parskip=0pt
\subsection{\textbf{Possible biases in constructing Barry}}

To examine the possibility of bias in results, let us consider our construction of Barry's brain's basic architecture.  We set Barry as a feed-forward network, with zero connectivity between nodes in a layer, and a constant back-propagation rule that permits no stochasticity into network evolution.  It is possible that Barry's true cognitive scheme is not as rigidly compartmentalized.  Further, we have enacted the learning by assuming a particular desire for efficiency: that it is output error that Barry seeks to minimize.  

Now, Barry's brain may be designed for efficiency at \textit{something}, but must it necessarily be minimizing output error?  Perhaps output error reduction is not Barry's agenda.  Perhaps instead, for example, Barry seeks to minimize the amount of work required to produce an answer, regardless of whether that answer is correct.  This is a possibility that would be strongly supported by any teacher or peer who shared a class with Barry during any time-frame within his grade school years (private communications, 1987-98).  

In short, other choices for constructing either the basic architecture or the learning algorithm might have yielded different outcomes (see \textit{Future Directions}).

\subsection{\textbf{Insight into the functionality of real biological networks}}

\vspace{50mm}

\subsection{\textbf{Future Directions}}

Let us now ruminate for a moment.  Let's imagine that we know absolutely the cost function to be minimized.  That is: We know Barry's aim in life.  Let us further imagine that we have developed a method to identify the global minimum of this cost function - and to calculate it within half a human lifetime.  In this case, the emergent network structure might contain clues regarding \textit{how} Nature designs a real central nervous system.   

What might this cost function be?  How might it relate to architectural complexity, robustness of certain activity patterns, variability of other patterns, redundancy, and other variables that a nervous system must consider when attempting to keep its host alive?  Further, is there just one cost function, or a system of cost functions?  

Unfortunately, to use the methodology devised in this paper to ascertain what cost function Barry is minimizing, we require a cost function to minimize.  The constraints we have erected are too rigid for such an exploration; Barry would require a bit of fleshing out.  Adding a feedback flow of information, for example, would be a sound place to begin.  Most people have memories, or at least we are under the impression that we do.

As publishable as this adventure sounds, however, we shall leave it in the hands of more wily souls.  We have obtained our result (87.2\% Yes!).  It is a rather delightful result and so we shall take it to be correct - and plan no future directions.
\end{multicols}
\begin{multicols}{2}
\noindent
\tiny{\section{MATERIALS AND METHODS}}
\small{
\subsection{\textbf{Feed-forward activation}}
We use a sigmoid activation function for the neurons because we assume that Barry is human to zeroth order and that a sigmoid function - which saturates at both ends of the input distribution - captures that feature.  The sigmoid y(x) is written:
\begin{align*}
  y(x) &= \sigma(w \cdot x + b), 
\end{align*}
where
\begin{align*}
  \sigma(z) &= \frac{1}{1 + e^{-z}}, 
\end{align*}
where each neuron receives i inputs $x_i$ and i weights $w_i$ and possesses a bias b.  For each layer $l$, then, one can compute output $a^l$ given the input ($z^l$):
\begin{align*}
  z^l &= w^la^{l-1} + b^l;\\
  a^l &= \sigma(z^l)
\end{align*}   

\subsection{\textbf{Back propagation}}

We use a quadratic cost function with a regularization term to penalize high weights.  A cross-entropy (logarithmic) cost function better approximates learning in most humans (e.g. Golik et al. 2013), but for Barry's case the quadratic form captures Barry's tendency to learn extremely slowly from large mistakes (Report Card: Algebra 1995).  The cost function $C(w,b)$ is written as: 
\begin{align*}
  C(w,b) &= \frac{1}{2n}\sum_j [y_j(x_j) - a_j^L]^2 + \frac{\lambda}{2n}\sum_{i,j} w_{ij}^2
\end{align*}
\noindent
where $w_{ij}$ are the synapse weights, b are the node biases, x are the training inputs, y(x) represents the desired output, $a^L$ is the activation of the final input (or, the ultimate output at layer L), $n$ is the number of training examples, and $\lambda$ is the regularization strength.  

A gradient descent is used to minimize the cost function.  This procedure entails a set of steps that iteratively relate the output error $\delta_j^l$ of layer $l$ to the updating of weights and biases of layers $l$ and ($l+1$) - beginning from the final layer $L$ and proceeding backwards.  The steps are standard:\\

\noindent
\textit{Back-propagation steps}
\begin{itemize}
  \item $\delta_j^L = \frac{\partial C}{\partial a_j^L}\sigma'(z_j^L)$, where $\sigma$' is the derivative of the activation function.
  \item Taking $\frac{\partial C}{\partial a_j^L} = (a_j^L - y_j)$ yields:\\ $\delta_j^L = (a_j^L - y_j)\sigma'(z_j^L)$. 
  \item To propagate this error, we define $\delta_j^l$ for each layer $l$: $\delta_j^l \equiv \frac{\partial C}{\partial z_j^l}$.
  \item We obtain: $\delta_j^l = \sigma'(z^l)\sum_j w^{l+1}_{ij}\delta_j^{l+1}$.
  \item We can relate the error $\delta_j^L$ to the quantities of interest - the weights $w_{ij}$ and biases $b_j$ - as follows:
  \begin{itemize}
    \item $\frac{\partial C}{\partial w_{jk}^l} = a_k^{l-1}\delta_j^l$;
    \item $\frac{\partial C}{\partial b_j^l} = \delta_j^l.$    \end{itemize}
\end{itemize}    
\noindent
  
\subsection{\textbf{Choices of user-defined parameters}}

The regularization term $\lambda$ was varied from zero (the limit in which high weight is irrelevant, or rather, is tamed only by the sigmoid activation) through 1 in increments of 0.01 (a regime of high penalty on large weights), and then from 1 to 100,000 in increments of 10 (a regime in which weight again becomes increasingly irrelevant).  This breadth of range was examined in order to explore the network's sensitivity to performance in a noisy environment. 

The step size for gradient descent (or \lq\lq learning rate\rq\rq) was set to 0.1, for no particular reason.

\subsection{\textbf{Training data}}

Question Set 1 contained 10,002 questions.  Question Set 2 contained 7,345 questions for the regime in which Barry had recently engaged in a fight with Janice, and 9,557 questions for the regime in which he had not; 2701 of these questions overlapped.  (\textit{For access to these supplementary materials, please email the author.})

Batches were made of 100 questions each.  Learning proceeded until the percent accuracy on validation tests asymptoted, or until it was determined that learning would in fact probably not occur at all (see \textit{Results}).

As these questions were originally posed in a high school hallway during passing time, these training sets are able to probe the network\rq s ability to perform in an extremely noisy environment.
}
\end{multicols}
\bibliographystyle{acm}
\nocite{*}
\bibliography{refs}

\end{document}